\documentclass[twocolumn]{pasj01}
\usepackage{ulem}

\begin{document} 
\Received{2017/12/07}
\Accepted{2018/03/14}

\title{Blue wing enhancement of the chromospheric Mg\emissiontype{II}~h~and~k lines in a solar flare}

\author{Akiko \textsc{Tei},\altaffilmark{1,}$^{*}$
	Takahito \textsc{Sakaue},\altaffilmark{1}
	Takenori J. \textsc{Okamoto},\altaffilmark{2}
	Tomoko \textsc{Kawate},\altaffilmark{3}
	Petr \textsc{Heinzel},\altaffilmark{4}
	Satoru \textsc{Ueno},\altaffilmark{1}
	Ayumi \textsc{Asai},\altaffilmark{1}
	Kiyoshi \textsc{Ichimoto},\altaffilmark{1}	and
	Kazunari \textsc{Shibata},\altaffilmark{1}}
\altaffiltext{1}{Kwasan and Hida Observatories, Kyoto University, Kyoto 607-8471, Japan}
\altaffiltext{2}{National Astronomical Observatory of Japan, Mitaka, Tokyo 181-8588, Japan}
\altaffiltext{3}{Institute of Space and Astronautical Science, Sagamihara, Kanagawa 252-5210, Japan}
\altaffiltext{4}{Astronomical Institute, Czech Academy of Sciences, 25165 Ond\v{r}ejov, Czech Republic}
\email{teiakiko@kwasan.kyoto-u.ac.jp}


\maketitle

\begin{abstract}

We performed coordinated observations of AR 12205, which produced a C-class flare on 2014 November 11, with the \textit{Interface Region Imaging Spectrograph} (\textit{IRIS}) and the Domeless Solar Telescope (DST) at Hida Observatory.
Using spectral data in the Si\emissiontype{IV}~1403~\AA, C\emissiontype{II}~1335~\AA, and Mg\emissiontype{II}~h~and~k lines from \textit{IRIS} and the Ca\emissiontype{II}~K, Ca\emissiontype{II}~8542~\AA, and H$\alpha$ lines from DST, we investigated a moving flare kernel during the flare.
In the Mg\emissiontype{II}~h line, the leading edge of the flare kernel showed the intensity enhancement in the blue wing, and the smaller intensity of the blue-side peak (h2v) than that of the red-side one (h2r).
The blueshift lasted for 9--48~s with a typical speed of 10.1~$\pm$~2.6~km~s$^{-1}$ and it was followed by the high intensity and the large redshift with a speed of up to 51~km~s$^{-1}$ detected in the Mg\emissiontype{II}~h line.
The large redshift was a common property for all six lines but the blueshift prior to it was found only in the Mg\emissiontype{II} lines.
A cloud modeling of the Mg\emissiontype{II}~h line suggests that the blue wing enhancement with such peak difference can be caused by a chromospheric-temperature (cool) upflow.
We discuss a scenario in which an upflow of cool plasma is lifted up by expanding hot plasma owing to the deep penetration of non-thermal electrons into the chromosphere.
Furthermore, we found that the blueshift persisted without any subsequent redshift in the leading edge of the flare kernel during its decaying phase.
The cause of such long-lasting blueshift is also discussed.
\end{abstract}

\section{Introduction}

Solar flares are the most energetic phenomena observed in the solar atmosphere, releasing a huge amount of energy up to 10$^{32}$ ergs on a typical timescale of minutes to tens of minutes.
Magnetic energy in the solar atmosphere is released through magnetic reconnection and converted to various kinds of energy: kinetic energy of mass ejection, thermal energy of hot plasma, and non-thermal energy of accelerated particles \citep{shibata11}.
On the basis of observations and theories, the standard flare reconnection model has been developed \citep{carmichael64, sturrock66, hirayama74, kopp76}.
From the viewpoint of energy transportation, it is unclear not only where and how the energy is released, transported, and converted to thermal and kinetic energy, but also how the lower atmosphere responds to the energy input.
In chromospheric lines such as the H$\alpha$ and Ca\emissiontype{II}~H~and~K, a pair of long and curved bright areas (flare ribbons) and strongly bright patches in flare ribbons (flare kernels) have been observed during flares \citep{svestka76}.
Such chromospheric flares are thought to be excited by the energy transferred from the corona into the chromosphere.

A number of authors have proposed mechanisms of the energy transport from the flare site in the corona to the chromosphere: e.g., electron beams \citep{brown73}, thermal conduction \citep{hirayama74}, soft X-ray (SXR) irradiation \citep{henoux78}, ion beams \citep{lin03}, and Alfv\'{e}n waves \citep{fletcher08}.
A beam of accelerated (non-thermal) electrons is a key mechanism that has been widely discussed, since strong enhancement of chromospheric (and occasionally photospheric) emission is temporally and spatially correlated with hard X-ray (HXR) emission and microwave radiation (e.g., \cite{asai12}) that are generated by the non-thermal bremsstrahlung \citep{brown71} and gyrosynchrotron emission \citep{takakura66}, respectively.

In relation to such a chromospheric heating, hot (8--25 MK) and dense upflow (up to 400~km~s$^{-1}$) along the flare loops has been detected in the wavelength range of SXR and extreme ultraviolet (EUV) during flares \citep{doschek80,antonucci82,strong84,milligan15}, which is called the chromospheric evaporation.
In the chromospheric evaporation scenario, high-energy (accelerated) particles go into and collide with the chromospheric plasma, which causes a rapid increase of temperature and pressure in the chromosphere \citep{hirayama74,fisher85}.
The heated plasma expands drastically upward along the magnetic field lines almost at its sound speed.
As a result, the plasma is observed as coronal hot loops in the SXR and EUV.
At the same time, the chromospheric plasma is strongly compressed and the downward propagating shock waves are excited in the chromosphere, because of the low chromospheric sound speed.
In this compressed region, radiative cooling becomes so effective that the temperature does not increase so much.
Consequently, the density increases more to keep the pressure balance, which leads to further effective radiation.
Therefore, the chromospheric condensation accompanying strong compression and downflow takes place.
Indeed, redshifts and enhancements in the red wing (red asymmetry) have been observed in the chromospheric lines in the impulsive phase of flares (e.g., \cite{svestka62, ichimoto84, shoji95, ding95}), which can be explained by condensation downflow in the chromosphere.
\citet{ichimoto84} measured redshifts and downward Doppler velocities in the H$\alpha$ line during flares quantitatively and established that downflow ($\sim$~50 km~s$^{-1}$) can be explained by the momentum balance with the chromospheric evaporation upflow.
This is confirmed by \citet{shoji95} that investigated other lines such as the Ca\emissiontype{II}~K and He D$_3$ as well as the H$\alpha$ line and showed that the velocity derived from the H$\alpha$ spectra was consistent with that from the He D$_3$ line (50--100~km~s$^{-1}$).

On the other hand, blue asymmetry in the chromospheric lines has also been reported \citep{svestka62, canfield90,heinzel94,kerr15}.
\citet{svestka62} studied the line asymmetry in 244 spectra of 92 individual flares observed in the H$\alpha$, Ca\emissiontype{II}~H~and~K, and He\emissiontype{I} lines.
They concluded that 80\% of the flares show red asymmetry and 23\% of the flares contain regions with blue asymmetry.
Furthermore, they showed that the blue asymmetry predominantly occurs in the early phase of the flares.
In \citet{heinzel94}, the Balmer and Ca\emissiontype{II}~H lines showed blue asymmetry at the same time during the onset phase of a flare.
Some researchers have attempted to explain the origin of the blue asymmetry \citep{canfield90,heinzel94,ding97,kuridze15,kuridze16}, but we have not obtained consensus.

The chromosphere in solar flares has been spectroscopically observed mostly from the ground so far.
The frequently-used lines have been the Balmer lines (mainly the H$\alpha$) and the Ca\emissiontype{II}~H,~K,~and 8542~\AA\ in visible and near infrared.
In addition to them, there are several lines in ultraviolet having strong emissions from the chromosphere that are not available from the ground.
The Mg\emissiontype{II} lines are the representative ones.
Now we can use the Mg\emissiontype{II}~h and k lines detected by the \textit{Interface Region Imaging Spectrograph} (\textit{IRIS}: \cite{depontieu14}), which was launched in 2013.
Furthermore, other chromospheric and transition region lines such as the C\emissiontype{II} and Si\emissiontype{IV} lines are also available with \textit{IRIS}.
With these ultraviolet lines, numerous studies of solar flares have been conducted.
\citet{kerr15} and \citet{liu15} studied the response of the Mg\emissiontype{II} lines to solar flares and presented the spatial and temporal behavior of the physical quantities.
\citet{kerr15} showed very intense, spatially localized energy input at the outer edge of the ribbon, resulting in redshifts equivalent to velocities of 15--26~km~s$^{-1}$, line broadenings, and line asymmetry in the most intense sources.
\citet{graham15} presented the dynamic evolution of chromospheric evaporation and condensation in a flare ribbon.
Successive brightenings in the impulsive phase displayed the same initial coronal upflows of up to 300~km~s$^{-1}$ and chromospheric downflows up to 40~km~s$^{-1}$.
See also \citet{li15}, \citet{li17}, and \citet{brosius17}, which investigated chromospheric evaporation and condensation using ultraviolet lines by \textit{IRIS}.
Recently, a narrow and dark moving feature has been observed at the front of flare footpoints in the He\emissiontype{I}~10830~\AA\ \citep{xu16}.
\cite{xu16} suggest that theoretically the dark feature in the He\emissiontype{I}~10830~\AA\ can be produced under special circumstances by non-thermal electron collisions or photoionization followed by recombination.
This might be related to the blue asymmetry observed in the chromospheric lines.

\begin{figure*}[htbp]
\begin{center}\FigureFile(160mm,80mm){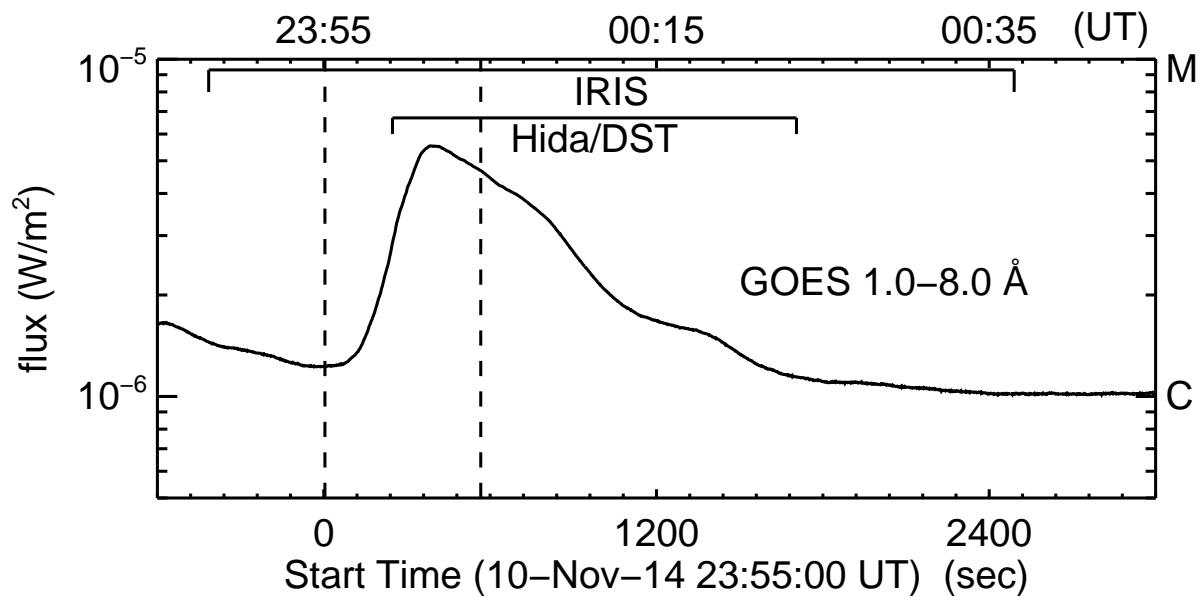}\end{center}
\caption{
\textit{GOES} soft X-ray (1.0--8.0~\AA) light curve of the C5.4 flare, which started at 23:55, peaked at 00:01, and ended at 00:08 (UT) on 2014 November 10 and 11.
The upper and lower horizontal axes are the time ``hh:mm" (UT) and the elapsed time from the start time of the flare, respectively.
Two dashed lines indicate the time range of figure~\ref{fig-lc_multi}.
Observation times of \textit{IRIS} and Hida DST are also shown.
}
\label{fig-goes}
\end{figure*}

In this paper, we report chromospheric dynamics related to a C5.4 flare on 2014 November 11 simultaneously observed by \textit{IRIS} and the Domeless Solar Telescope (DST: \cite{nakai85}) at Hida Observatory of Kyoto University.
We show the temporal and spatial evolution of six chromospheric and transition region lines in relation to a moving flare kernel that was observed during the impulsive phase of the flare.
In section 2, we describe the data set and data reduction.
Then, observational description of the flare kernel and our method to analyze the data are shown in section 3.
Section 4 shows the results.
Finally, we discuss the interpretation of observational results and present the summary in section 5.

\begin{figure*}[htbp]
\begin{center}\FigureFile(160mm,220mm){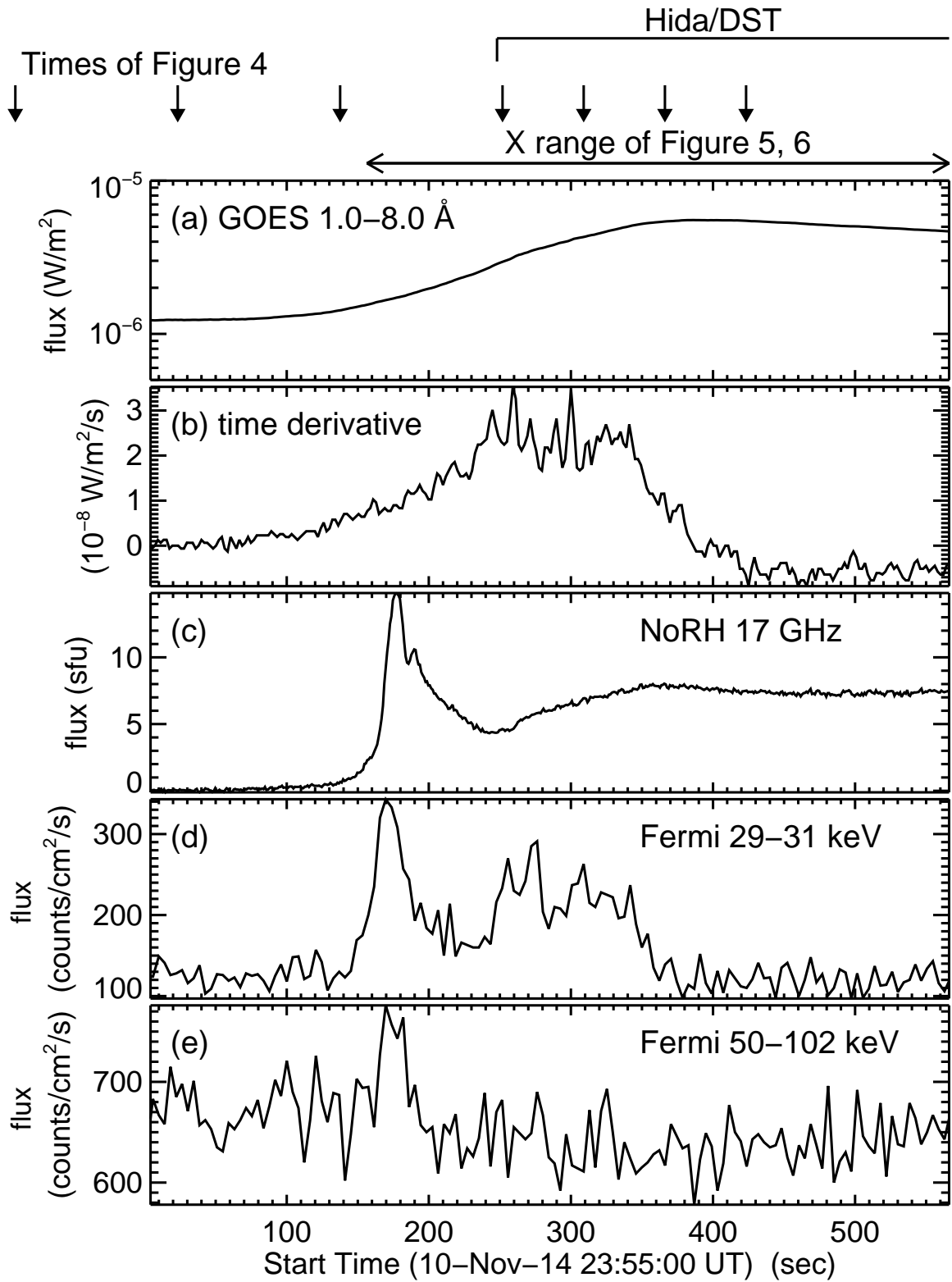}\end{center}
\caption{
Temporal variation of high energy emissions.
(a) \textit{GOES} soft X-ray flux of the 1.0--8.0~\AA\ channel. 
(b) The time derivative of the \textit{GOES} soft X-ray flux.
(c) NoRH 17~GHz microwave from the flare region. The preflare flux is subtracted.
(d-e) \textit{Fermi}/GBM hard X-ray flux in the range of 29--31~keV and 50--100~keV, respectively.
The horizontal axis is the elapsed time from the start time of the flare.
The times of the seven snapshots in figure~\ref{fig-ker} are indicated by the arrows.
The double-headed arrow denotes the X (horizontal) range of figure~\ref{fig-tymap_iris} and figure~\ref{fig-tymap_dst}.
The observation time of Hida DST is also shown.
}
\label{fig-lc_multi}
\end{figure*}

\begin{figure*}[htbp]
\begin{center}\FigureFile(160mm,120mm){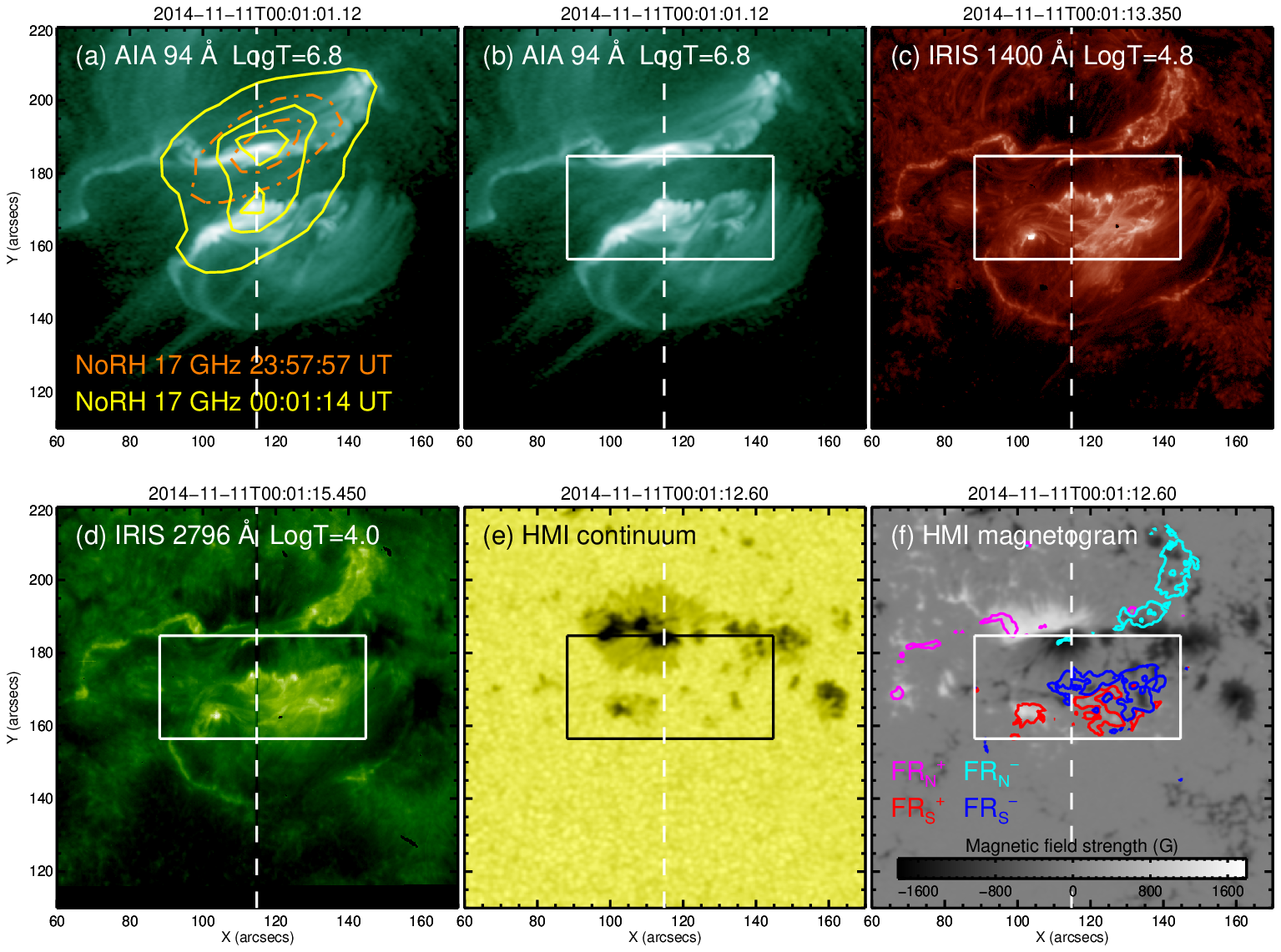}\end{center}
\caption{
Snapshots of the flare region around the peak time of the flare at 00:01~UT.
(a-b) \textit{SDO}/AIA 94~\AA.
(c) \textit{IRIS} SJI 1400~\AA.
(d) \textit{IRIS} SJI 2796~\AA.
(e) \textit{SDO}/HMI continuum.
(f) \textit{SDO}/HMI line-of-sight magnetogram.
Characteristic temperatures are shown in (a)-(d).
In panel (a), we overlaid contour images of NoRH 17~GHz: orange dash-dotted lines show 40\% and 80\% levels of the peak intensity at 23:57:57~UT and yellow solid lines show 40\%, 80\%, and 94\% levels of the peak intensity at 00:01:14~UT.
In panel (f), two sets of two conjugate flare ribbons are shown by the overlaid contour image (10\% level of the peak intensity of the \textit{IRIS} SJI 2796~\AA): the northern flare ribbon with positive and negative polarity (FR$_{\rm N}^+$ and FR$_{\rm N}^-$) by magenta and cyan line, and the southern flare ribbon with positive and negative polarity (FR$_{\rm S}^+$ and FR$_{\rm S}^-$) by red and blue line, respectively.
The vertical dashed line is the location of the \textit{IRIS} slit.
The white box is the region shown in figure~\ref{fig-ker}.
(Color online)}
\label{fig-fov}
\end{figure*}

\section{The data set and data reduction}

We performed coordinated observations (HOP 275) of NOAA AR 12205 located near the disk center (N13$^\circ$ W17$^\circ$) from 2014 November 10 to 11 with \textit{IRIS} and DST.
The active region produced a C5.4 flare that started at 23:55 UT on November 10, peaked at 00:01 UT, and ended at 00:08 UT on November 11 (figure~\ref{fig-goes}).

\textit{IRIS} observed the flare at the fixed slit location (sit-and-stare mode) from 23:48:05 UT on November 10 to 00:36:30 UT on November 11 (observation time is shown in figure~\ref{fig-goes}).
The field of view (FOV) by the slit-jaw imager (SJI) was 120\arcsec$\times$162\arcsec\ with a sampling of 0\farcs166~pixel$^{-1}$.
The spatial sampling along the \textit{IRIS} slit by the spectrograph was also 0\farcs166~pixel$^{-1}$.
The temporal cadence was 9.5~s for the spectral data and 29~s for the SJI filtergrams.
In this study, we used the Mg\emissiontype{II}~h~2803~\AA, Mg\emissiontype{II}~k~2796~\AA, C\emissiontype{II}~1335~\AA, and Si\emissiontype{IV}~1403~\AA\ lines, which have the formation temperatures of ${\rm Log}~T~{\rm [K]}= $ 4.0, 4.0, 4.3, and 4.8, respectively \citep{depontieu14}.
We used mainly the Mg\emissiontype{II}~h line rather than the Mg\emissiontype{II}~k line, because another line locates in the blue wing of the Mg\emissiontype{II}~k line and this causes difficulties of correct measurements of characteristic parameters of the line as described later.
The Mg\emissiontype{II}~k line behaves similarly to and has relatively higher intensity than the Mg\emissiontype{II}~h line.
We note that the signal from the Fe\emissiontype{XXI} line, which can be used for the indication of the coronal evaporation upflow, was too weak to be detected.

DST observed the flare from 23:59:08 UT on November 10 to 00:23:25 UT on November 11 (as shown in figure~\ref{fig-goes}).
Raster-scanned maps by the slit that moved from west to east on the solar surface were obtained with the horizontal spectrograph on DST with a repeat cadence of 10~s.
We used the spectral data of the Ca\emissiontype{II}~K 3934~\AA, Ca\emissiontype{II}~8542~\AA, and H$\alpha$ 6563~\AA\ lines, which all form at around the chromospheric temperature ($\sim 10^4\ {\rm K}$).
The data were taken by three identical cameras: Prosilica GE1650, and the timing to get the data were synchronized in all the cameras.
The spatial samplings in the vertical direction to the slit were 0\farcs64~pixel$^{-1}$ and those along the slit were 0\farcs27~pixel$^{-1}$, 0\farcs28~pixel$^{-1}$, and 0\farcs36~pixel$^{-1}$ for the Ca\emissiontype{II}~K, Ca\emissiontype{II}~8542~\AA, and H$\alpha$ lines, respectively.
We note that the actual spatial resolution was about 2\arcsec\ due to the seeing effect.
The spectral sampling and the velocity that corresponds to the spectral sampling were 15~m\AA~pixel$^{-1}$ and 1.1~km~s$^{-1}$ for the Ca\emissiontype{II}~K line, 15~m\AA~pixel$^{-1}$ and 0.5~km~s$^{-1}$ for the Ca\emissiontype{II}~8542~\AA\ line, and 22~m\AA~pixel$^{-1}$ and 1.0~km~s$^{-1}$ for the H$\alpha$ line.
We used the spectral data smoothed over $\pm$5, 15, and 10 data points along the wavelength for the Ca\emissiontype{II}~K, Ca\emissiontype{II}~8542~\AA, and H$\alpha$ line, respectively because physical parameters cannot be derived due to the low signal-to-noise ratio of them.
This smoothing does not affect the measurements of velocities described in the section 3.
The smoothing leads an artificial broadening of profiles, but the absolute values of line widths are not important in this study.

Time profiles of the high-energy emissions are obtained by the Nobeyama Radioheliograph (NoRH: \cite{nakajima94}) and the \textit{Fermi}/Gamma-ray Burst Monitor (GBM: \cite{meegan09}).
The NoRH data were created by SSW function ``norh\_tb2flux", with the option ``abox~$=$~[0, 220, 80, 330]" (with the unit [arcsecs]) covering the flare region.
The \textit{Fermi}/GBM n0 detector was best facing the Sun during this flare and used in this study.

We also used 94~\AA, 304~\AA, and 1600~\AA\ images of the Atmospheric Imaging Assembly (AIA: \cite{lemen12}) and continuum images and line-of-sight magnetograms of the Helioseismic and Magnetic Imager (HMI: \cite{scherrer12}).
Both AIA and HMI are installed in the \textit{Solar Dynamics Observatory} (\textit{SDO}: \cite{pesnell12}).

We conducted co-alignments between the data of the different instruments, using images and maps taken around 23:59:08~UT.
First, we aligned the intensity map at the H$\alpha$ $-$8~\AA\ from DST to the continuum image of HMI.
Second, maps in the Ca\emissiontype{II}~K $-$5.5~\AA\ and Ca\emissiontype{II}~8542~\AA\ $-$5.0~\AA\ from DST were aligned to the H$\alpha$ map.
Third, the AIA 1600~\AA\ image and the \textit{IRIS} Si\emissiontype{IV} image were used for alignment between all \textit{SDO} images and \textit{IRIS} data.
Combining these procedures, we can finally remove offsets between images and maps.

\begin{figure*}[]
\begin{center}\FigureFile(160mm,140mm){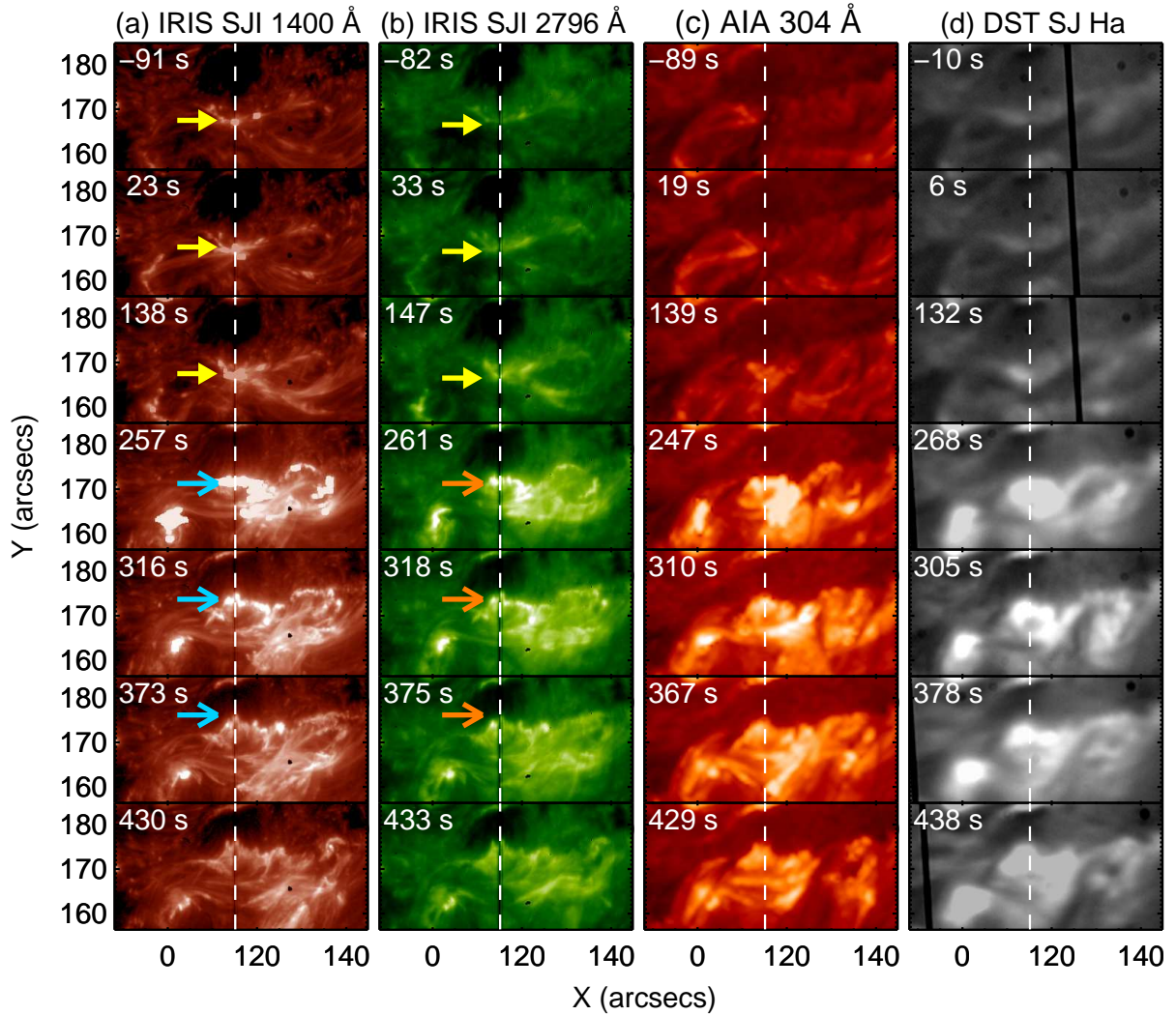}\end{center}
\caption{
Time series of the flare region in the white box of figure~\ref{fig-fov}.
In each panel, the elapsed time from the start time of the flare at 23:55~UT ($t$) is shown.
The dashed line indicates the \textit{IRIS} slit location.
In (a) and (b), the bright feature staying at the same slit position is indicated by the yellow arrows.
The (a) cyan and (b) orange arrows indicate the position of the moving flare kernel.
(Color online)}
\label{fig-ker}
\end{figure*}

\begin{figure*}[htbp]
\begin{center}\FigureFile(160mm,120mm){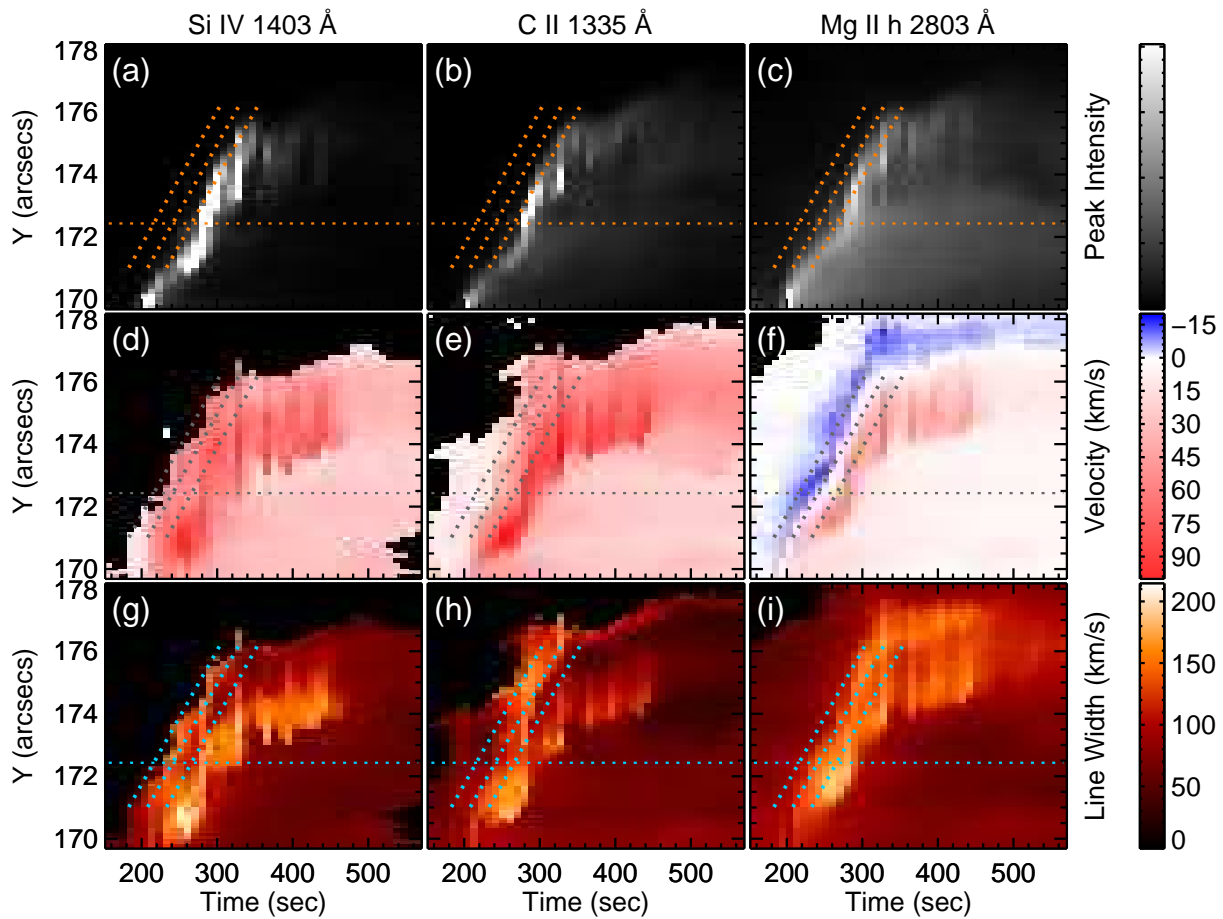}\end{center}
\caption{
Space--time plots of (a-c) line peak intensity, (d-f) Doppler velocity, and (g-i) line width of the \textit{IRIS} lines.
Three diagonal dotted lines are drawn at the same space and time on all the panels to compare among different panels and 
each slope corresponds to the apparent velocity of 31~km~s$^{-1}$.
The detail of the location indicated by the horizontal dotted line ($y_{\rm slit}=$172\farcs 4) is shown in figure~\ref{fig-ivw_iris}.
The black areas in the panel (d-i) present the regions where physical parameters cannot be derived due to too weak intensity.
We also show color bars for the three quantities on the right hand side.
Intensity plots are shown in logarithmic scale and the lighter color shows the brighter value.
(Color online)}
\label{fig-tymap_iris}
\end{figure*}

\begin{figure*}[htbp]
\begin{center}\FigureFile(160mm,120mm){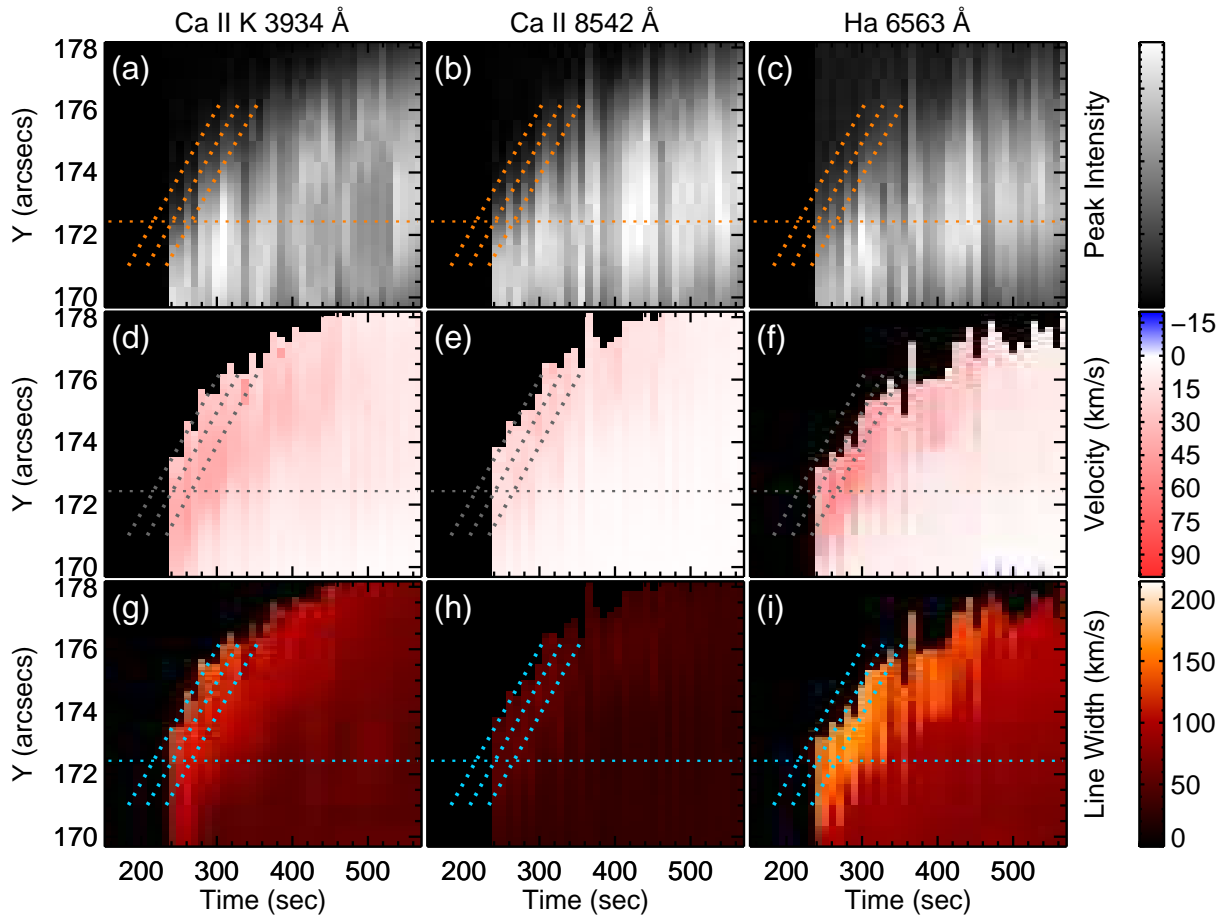}\end{center}
\caption{
Same as figure~\ref{fig-tymap_iris}, but for the DST lines.
Three diagonal dotted lines are drawn at the same space and time as those in figure~\ref{fig-tymap_iris}.
The detail of the location indicated by the horizontal dotted line ($y_{\rm slit}=$172\farcs 4) is shown in figure~\ref{fig-ivw_dst}.
Note that the DST observation started at $t$ = 242~s.
(Color online)}
\label{fig-tymap_dst}
\end{figure*}

\begin{figure*}[htbp]
\begin{center}\FigureFile(160mm,160mm){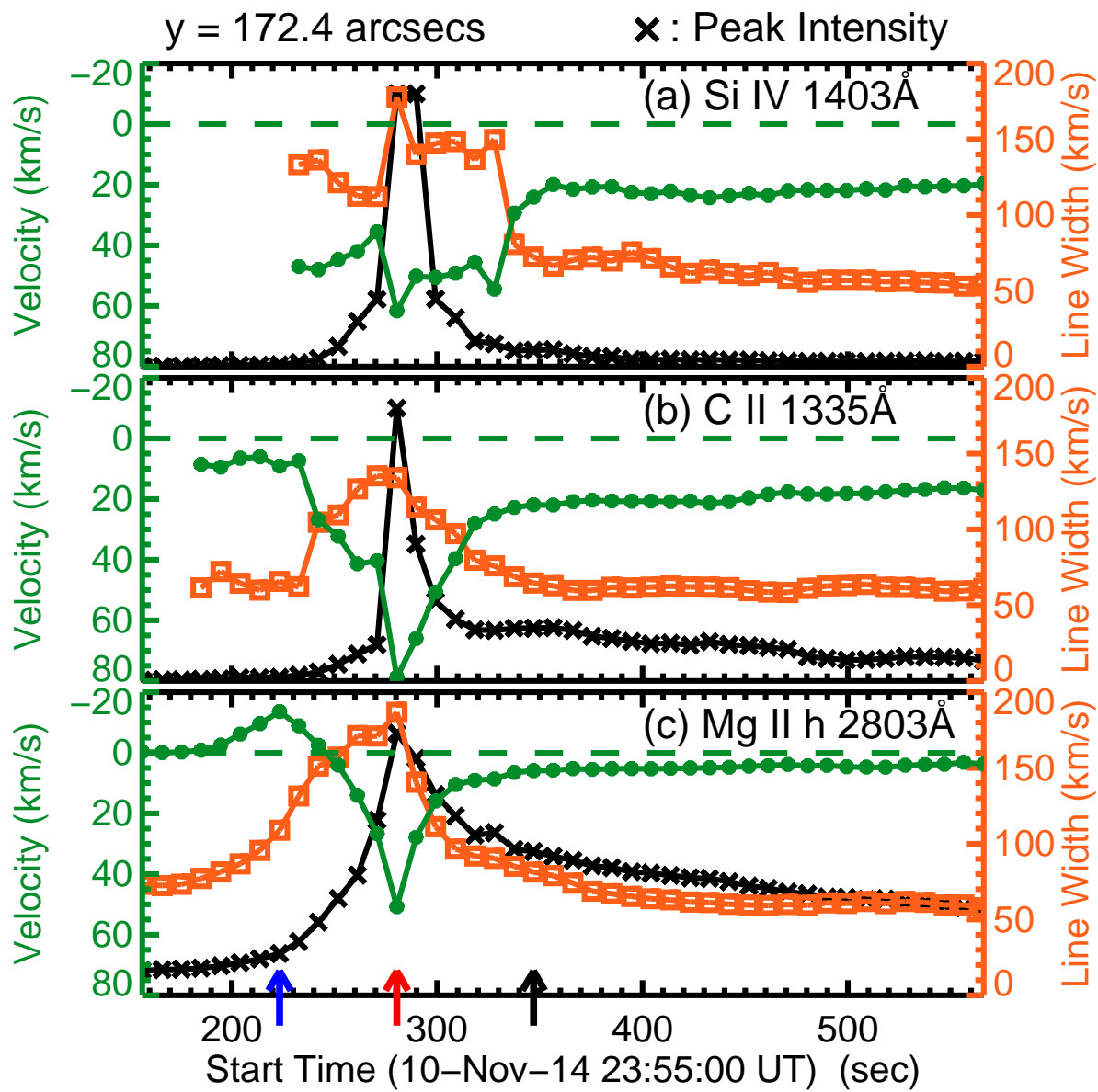}\end{center}
\caption{
Temporal evolution of the peak intensity (black cross), the Doppler velocity (green circle), and the line width (orange square) in the \textit{IRIS} lines at $y_{\rm slit}=~172\farcs 4$ on the \textit{IRIS} slit.
Vertical axes of peak intensities are arbitrary.
Negative velocity represents blueshift, while positive velocity redshift. 
Line width and velocity are not shown in the time period when the intensity was too weak to determine.
Three arrows at the bottom indicate the same times when the spectra shown in the panels in figure~\ref{fig-sp_iris} were taken.
The blue and red arrows are the time of the minimum velocity (blueshift) ($t$ = 223~s) and the maximum velocity (redshift) ($t$ = 280~s) in the Mg\emissiontype{II}~h line, respectively.
(Color online)}
\label{fig-ivw_iris}
\end{figure*}

\begin{figure*}[htbp]
\begin{center}\FigureFile(160mm,160mm){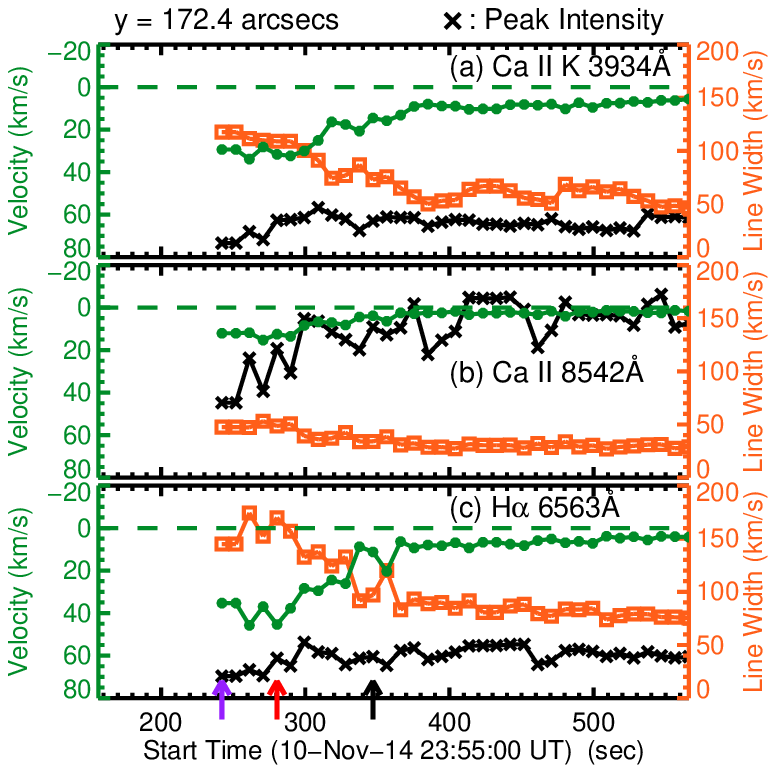}\end{center}
\caption{
The same as figure~\ref{fig-ivw_iris}, but for the DST lines.
Three arrows indicate the times when spectra shown in the panels in figure~\ref{fig-sp_dst} were taken.
The purple arrow is the first time of the DST observation ($t$ = 242~s).
The red arrow is the time of the maximum velocity (redshift) in the Mg\emissiontype{II}~h line, which is the same as the red arrow in figure~\ref{fig-ivw_iris}.
(Color online)}
\label{fig-ivw_dst}
\end{figure*}

\section{The flare kernel and its analysis}

In order to investigate the energy input and the dynamics, we looked at a flare kernel that appeared during the impulsive phase of the flare.
Here is the description of the flare kernel on which we focus in this paper.
We define two variables $t$ and $y_{\rm slit}$ as the elapsed time from the flare onset at 23:55:00~UT and the $y$ position along the \textit{IRIS} slit in the solar coordinates, respectively.
Figure~\ref{fig-lc_multi} shows the temporal variation of high energy emissions around the impulsive phase ($t=$150--360~s), which indicates the energy input from the corona to the lower atmosphere.
The impulsive phase consists of two bursts.
The first one was involved in the NoRH 17~GHz and \textit{Fermi} HXR flux during $t=$150--200~s.
During the second burst ($t=$200--360~s), the time derivative of the \textit{GOES} SXR and \textit{Fermi} HXR flux increased.
Figure~\ref{fig-fov} shows images of the flare region at the flare peak time.
There were two sets of two conjugate flare ribbons.
Each set of them was at the northern and southern part of the field-of-view (FOV) of figure~\ref{fig-fov}.
The northern flare ribbons (FR$_{\rm N}^+$ and FR$_{\rm N}^-$) were at two sides of the magnetic polarity inversion line of the delta-type sunspot.
The southern flare ribbons (FR$_{\rm S}^+$ and FR$_{\rm S}^-$) were involved in the mixed polarity region to the south of the delta-type sunspot.
The flare kernel that we focus on was located in FR$_{\rm S}^-$ (on the \textit{IRIS} slit ($y_{\rm slit}\sim$174\arcsec\ in figure~\ref{fig-fov}(c-d))).
As shown in figure~\ref{fig-ker}, the intensity of the flare ribbon was weak during the first burst and became much higher at $t\sim$200~s.
Note that the increase of the NoRH 17~GHz in the first burst ($t=$150--200~s in figure~\ref{fig-lc_multi}(c)) was involved in the northern flare (see figure~\ref{fig-fov}(a)), which is out of our scope.
During the second burst ($t=$200--360~s), the flare kernel moved $\sim$~9\arcsec\ northward along the slit with an apparent speed of 31~km~s$^{-1}$.
At $t\sim$360~s, the feature stopped moving in the vicinity of the penumbra and became less bright.

We investigated the temporal and spatial variation of dynamics around the flare kernel by measuring three quantities of the spectral lines: peak intensity, line width, and Doppler velocity as well as first moment velocity (intensity-weighted velocity across the line).
The original line profiles were used to derive the quantities from the \textit{IRIS} data (mostly emission profiles), while the DST line profiles (mostly absorption profiles) were processed after subtracting the reference profile from the original ones.
The reference profiles of the DST lines are the averaged spectra of a non-flaring region.
The rest wavelengths were defined by the central wavelengths of averaged spectra in non-flaring regions, which are expected to have intrinsic velocity of zero or less than a few km~s$^{-1}$.
Doppler velocity and line width were derived by the bisector method: the central position and the difference of the two wavelengths in red and blue sides of the line profile at which the intensity gets 30\% of the peak, respectively.
We adopted this intensity level since intensity enhancement in line wings, which will be focused on in this study, is well captured at this level (30\% level was also used by \cite{graham15}).
We derived Doppler velocity and line width with other intensity levels (at every 1\% from 20\% to 40\%) and confirmed that the values differ from those of the 30\% level within 5~km~s$^{-1}$ for Doppler velocity and 30~km~s$^{-1}$ for line width, respectively.
The instrumental width of the \textit{IRIS} FUV band (including the C\emissiontype{II} and Si\emissiontype{IV} lines)
and the NUV band (including the Mg\emissiontype{II} lines) is 25.85~m\AA\ and 50.54~m\AA, respectively \citep{depontieu14}, which are essentially negligible in this study.
Note that the line widths derived in this paper (the full width at 30\% maximum) are larger than those by the usual definition (the full width at half maximum (FWHM)).

\section{Results}

Figures~\ref{fig-tymap_iris} and \ref{fig-tymap_dst} are the space-time plots of the peak intensity, the Doppler velocity, and the line width along the \textit{IRIS} slit location for the \textit{IRIS} and DST lines, respectively, in the range to focus on the moving flare kernel.
We do not show the first moment velocity in this paper, since they evolved similarly with, but slightly smaller than, the Doppler velocity determined by the bisector method in all the lines.
The peak intensities of the \textit{IRIS} lines evolved similar to each other, while the brightening in the three DST lines persisted longer than in the \textit{IRIS} lines.
On the other hand, the temporal and spatial variations of line shifts look different: only in the Mg\emissiontype{II}~h line, the blueshift was confirmed.
No clear blueshift was observed in other five lines.
The blueshift appeared when the flare kernel started moving; the blueshift lasted during the second burst ($t=$200--360~s).
While the bluesift from 171\farcs 0 to 176\farcs 2 in $y_{\rm slit}$ was followed by the redshift in the Mg\emissiontype{II}~h line, the blueshift from 176\farcs 2 to 178\farcs 1 in $y_{\rm slit}$ persisted and was not followed by any clear redshift.
In the DST lines, the spatial structure was extended in the spatial direction compared to that in the \textit{IRIS} lines due to the seeing effect ($\sim$~2\arcsec).

Here we focus on the evolution of the peak intensity, the Doppler velocity, and the line width from 171\farcs 0 to 176\farcs 2 in $y_{\rm slit}$.
(The evolution from 176\farcs 2 to 178\farcs 1 in $y_{\rm slit}$ is described in the last paragraph of this section.)
The temporal variations of the three quantities at the pixels in the $y_{\rm slit}$ range were similar to each other, while redshifts continued longer in the positions from 174\farcs 0\ to 176\farcs 2\ in $y_{\rm slit}$.
The spatial width of the region with the blueshift in the Mg\emissiontype{II}~h line was $\sim$ 700 km in the traveling direction.
Recently \cite{xu16} observed the dark feature in the He\emissiontype{I}~10830~\AA\ line at the leading edge of flare footpoints before the burst phase.
The spatial width of the dark feature in their study (340--510~km) was smaller than that of the feature with blueshift in this study, but both occurred at the leading edge of flare footpoints.

With figures~\ref{fig-ivw_iris}~and~\ref{fig-ivw_dst}, we take a closer look at the typical temporal variation of the three quantities at a pixel: the peak intensity, the line width, and the Doppler velocity at $y_{\rm slit}=$172\farcs 4 (horizontal dotted line in figure~\ref{fig-tymap_iris}~and~\ref{fig-tymap_dst}).
At this location, the blueshift in the Mg\emissiontype{II}~h line was the largest among all the pixels from 171\farcs 0 to 176\farcs 2 in $y_{\rm slit}$.
The blueshift that was larger than 5~km~s$^{-1}$ lasted for 29~s with the maximum velocity of 14~km~s$^{-1}$ (shown by the blue arrow in figure~\ref{fig-ivw_iris}).
During this period, the intensity of the Mg\emissiontype{II}~h was still small but just started getting large.
The line width became broader when the blueshift appeared in the Mg\emissiontype{II}~h line.
After that, the large redshift in the Si\emissiontype{IV}, C\emissiontype{II}, and Mg\emissiontype{II}~h lines lasted for 67~s, 29~s, and 10~s (e-folding time) with the maximum velocity of 62~km~s$^{-1}$, 79~km~s$^{-1}$, and 51~km~s$^{-1}$, respectively.
The intensity and the line width also increased together.
The three quantities reached their maxima almost at the same time as indicated by the red arrow, while the line width in the C\emissiontype{II} line peaked 9.5~s earlier.
As for the DST lines, when the observation started (purple arrow in figure~\ref{fig-ivw_dst}), the blueshift in the Mg\emissiontype{II}~h line was almost over and all the DST lines showed the redshifts.
At $y_{\rm slit}=$172\farcs 4, the maximum velocity was 34~km~s$^{-1}$, 15~km~s$^{-1}$, and 46~km~s$^{-1}$ for the Ca\emissiontype{II}~K, Ca\emissiontype{II}~8542~\AA, and H$\alpha$ lines, respectively.
Then, the Doppler velocities decreased in the three lines.
The line widths varied similarly to the line shifts.
The brightening in the three DST lines peaked a little later and persisted longer than in the \textit{IRIS} lines.

\begin{figure*}[htbp]
\begin{center}\FigureFile(120mm,180mm){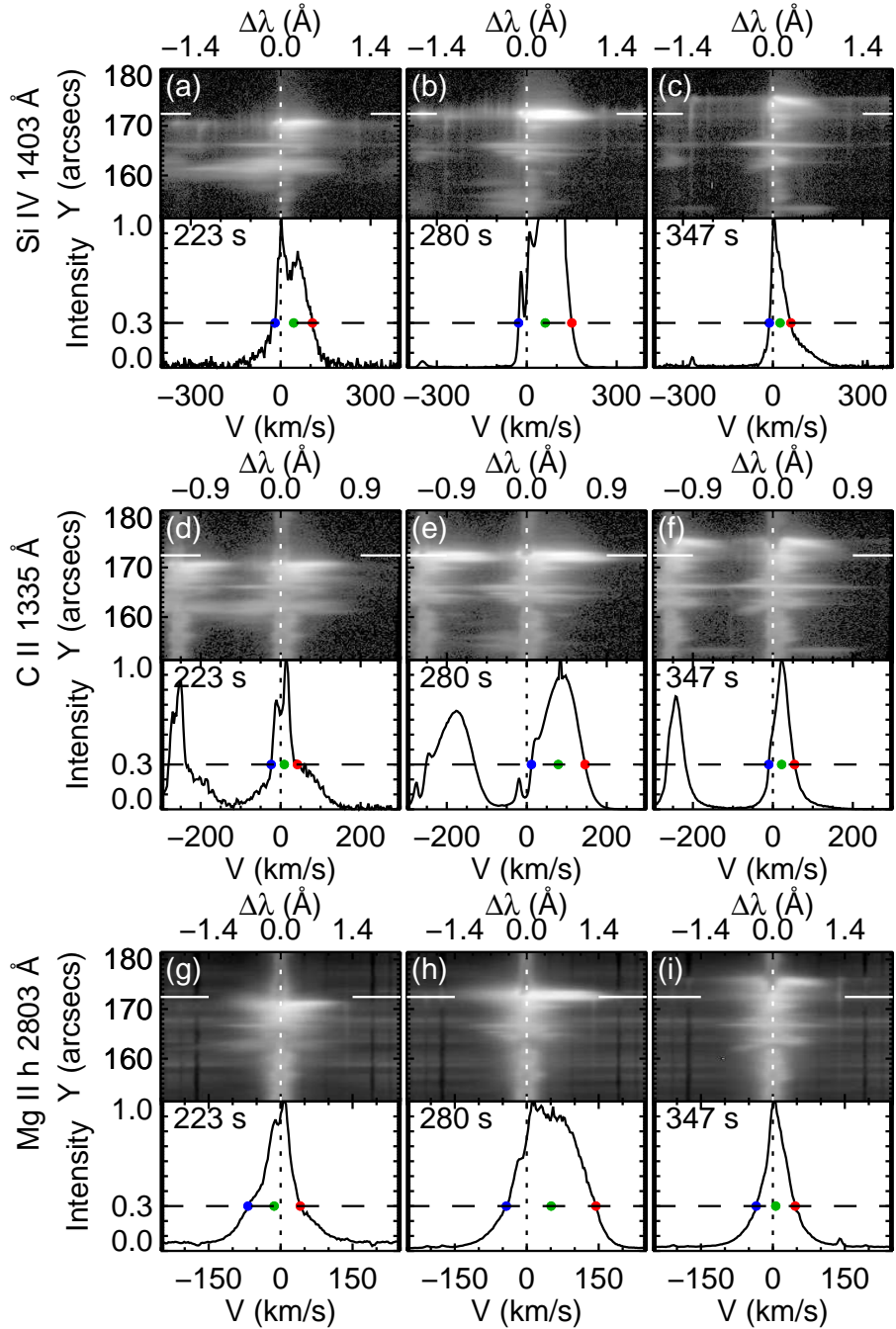}\end{center}
\caption{
Spectral images and profiles of the \textit{IRIS} (a-c) Si\emissiontype{IV}, (d-f) C\emissiontype{II}, and (g-i) Mg\emissiontype{II}~h at $t=$~223~s, $t=$~280~s, and $t=$~347~s, corresponding to the times of three arrows in figure~\ref{fig-ivw_iris}, respectively.
In the upper portion in each panel, the space--wavelength plot in the vicinity of the flare kernel is shown.
The original line profile at the spatial position $y_{\rm slit}=$~172\farcs4 (indicated by the horizontal solid lines in the upper image) is shown in the lower portion of each panel.
The peak intensity is normalized to unity and the horizontal axis is converted to the velocity from the wavelength.
The threshold of the intensity to define the bisector (30\% peak intensity) (dashed line) and the rest wavelength (dotted line) are shown.
The blue and red dots indicate the positions of the two wing wavelengths at 30\% peak intensity and the green dots show the bisector wavelength of the two wings.
Note that the intensity is saturated in panel (b) and the another C\emissiontype{II} line appears in panels (d-f), in the shorter wavelength of the C\emissiontype{II}~1335\ \AA.
(Color online)}
\label{fig-sp_iris}
\end{figure*}

\begin{figure*}[htbp]
\begin{center}\FigureFile(120mm,180mm){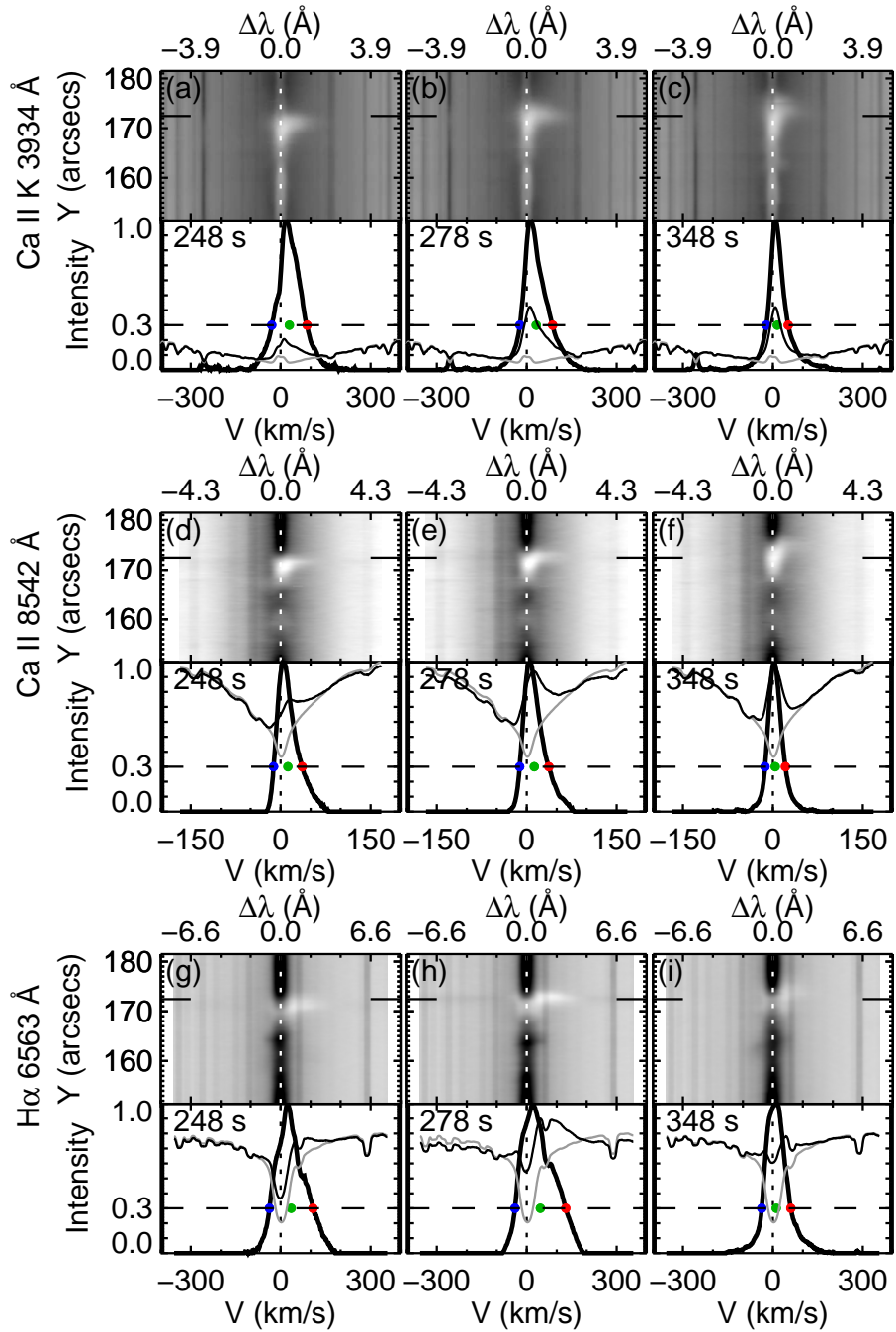}\end{center}
\caption{
Spectral images and profiles of the DST (a-c) Ca\emissiontype{II}~K, (d-f) Ca\emissiontype{II}~8542~\AA, and (g-i) H$\alpha$ at $t=$~248~s, $t=$~278~s, and $t=$~348~s, corresponding to the nearest times of three arrows in figure~\ref{fig-ivw_dst}, respectively.
In the upper portion in each panel, the space--wavelength plot in the vicinity of the flare kernel is shown.
The lower portion of each panel shows the original line profile (thin black line), the reference line profile (thin gray line), and the difference profile obtained by subtracting the reference profile from the original one (thick black line) at the spatial position $y_{\rm slit}=$~172\farcs4 (indicated by the horizontal solid lines in the upper region).
The peak intensity of the difference profile is normalized to unity in each panel.
Horizontal dotted line and colored dots are the same as those in figure~\ref{fig-sp_iris} but using the difference profile.
Note that the intensity of the original and reference line profiles are shown in arbitrary scale.
(Color online)}
\label{fig-sp_dst}
\end{figure*}

\begin{figure*}[htbp]
\begin{center}\FigureFile(160mm,200mm){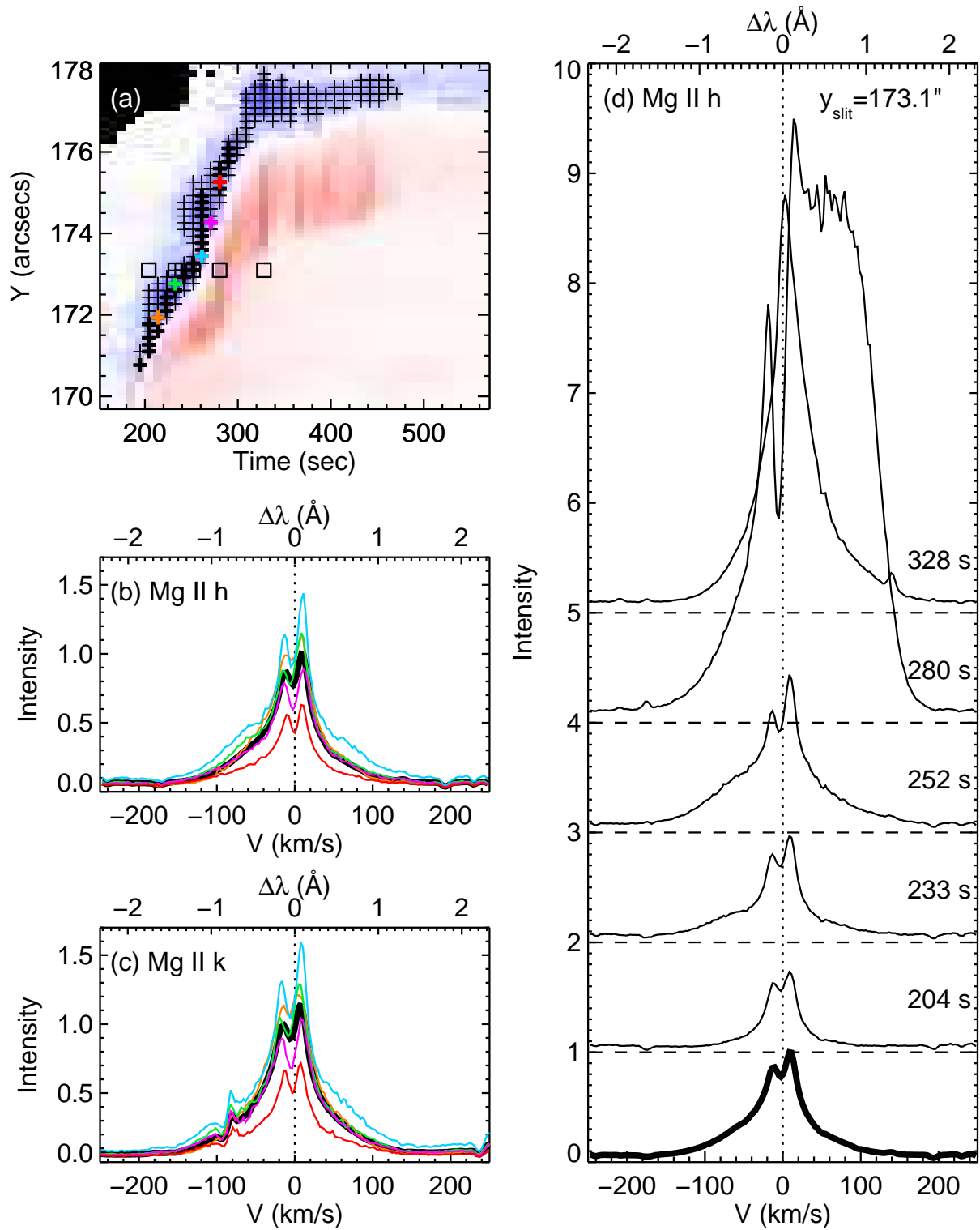}\end{center}
\caption{
The detail of the blueshift in the Mg\emissiontype{II} lines at the location from 171\farcs 0 to 176\farcs 2 in $y_{\rm slit}$, where the blueshift preceded the redshift.
(a) Distribution of the blueshift in the Mg\emissiontype{II}~h line.
Background is the same as in figure~\ref{fig-tymap_iris}(f).
Plus signs indicate the time and location at which the blueshift of the bisector was larger than 5~km~s$^{-1}$.
Each thick plus sign shows the time when the blueshift got maximum at each location.
Mg\emissiontype{II}~h~and~k line profiles at the positions of colored thick plus signs are shown in panels (b) and (c), respectively.
Each square sign indicates the time and location of each Mg\emissiontype{II}~h line profile in panel (d).
(b-c) Thick black line shows the average of the Mg\emissiontype{II} (b) h and (c) k line profiles at the positions indicated by thick plus signs in panel (a), respectively.
(d) An example of the temporal evolution of the Mg\emissiontype{II}~h line profiles at the positions indicated by the square signs in panel (a) (y$_{\rm slit}$=173\farcs 1).
Time passes from bottom up and the times are shown except the lowest profile, which is the same as the averaged one in panel (b).
In panels (b-d), the vertical dotted lines show the rest wavelength and peak intensity of the averaged line profile of Mg\emissiontype{II}~h is normalized to unity.
(Color online)}
\label{fig-peak}
\end{figure*}

The line profiles at the times indicated by the three arrows in figure~\ref{fig-ivw_iris} are shown in figure~\ref{fig-sp_iris}.
From figures~\ref{fig-ivw_iris}(c) and \ref{fig-sp_iris}(g-i), we can recognize that the blueshift in the Mg\emissiontype{II}~h line was due to the weak enhancement in the blue wing with the moderate peak intensity and that the redshift was due to the strong enhancement in the red wing with the large peak intensity.
In addition, as shown in figure~\ref{fig-sp_iris}(g), the blue-side peak (h2v) was smaller than the red-side one (h2r).
The Si\emissiontype{IV} and C\emissiontype{II} lines (figure~\ref{fig-sp_iris}(a-c) and (d-f)) show strong enhancement in the red wing, in particular at $t=$223 and 280~s, respectively.
Note that the profile shown in figure~\ref{fig-sp_iris}(b) was saturated.
Figure~\ref{fig-sp_dst} shows the similar plots as figure~\ref{fig-sp_iris} but for the DST lines at the times indicated by the three arrows in figure~\ref{fig-ivw_dst}, with the reference and original profiles.
In all the DST lines at $t=$248 and 278~s, no blueshift was confirmed and the strong emission especially in the red wing can be seen.
These characteristics were similar in all positions from $171\farcs 0$\ to $176\farcs 2$\ in $y_{\rm slit}$.
Figure~\ref{fig-peak} shows the detail of the blueshift in the Mg\emissiontype{II} lines from 171\farcs 0 to 176\farcs 2 in $y_{\rm slit}$.
In figure~\ref{fig-peak}(a), black plus signs show the distribution of the blueshift that was larger than 5~km~s$^{-1}$ in the location from 171\farcs 0\ to 178\farcs 1\ in $y_{\rm slit}$. 
Each of thick plus signs among them indicates the time of the peak blueshift at each pixel of the location from 171\farcs 0\ to 176\farcs 2\ in $y_{\rm slit}$, where the blueshift preceded the redshift.
Thick and thin lines in figure~\ref{fig-peak}(b-c) show the average and five examples of the Mg\emissiontype{II}~h~and~k line profiles at the positions indicated by black and colored thick plus signs in figure~\ref{fig-peak}(a), respectively.
As is the case with the profile in figure~\ref{fig-sp_iris}(g), the averaged profile of the Mg\emissiontype{II}~h line had the enhanced blue wing and the small intensity of the blue-side peak (h2v) compared to the red-side peak (h2r), which was the typical shape among the blueshifted line profiles.
The Mg\emissiontype{II}~k line behaved similarly with the Mg\emissiontype{II}~h (figure~\ref{fig-peak}(b-c)) while its intensity is larger and the blue wing is blended by another line.
In the Mg\emissiontype{II}~h line, the mean velocity of the blueshift indicated by the thick plus signs in figure~\ref{fig-peak}(a) was 10.1~km~s$^{-1}$ with the standard deviation of 2.6~km~s$^{-1}$.
The mean duration of the blueshift ($>$~5~km~s$^{-1}$) at one pixel location was 28~s with the minimum and maximum duration of 9~s and 48~s, respectively.
Figure~\ref{fig-peak}(d) shows an example of the temporal evolution of the Mg\emissiontype{II}~h line profiles at a specific location ($y_{\rm slit}$=173\farcs 1; square signs in figure~\ref{fig-peak}(a)): the transition from the blueshifted profiles to the redshifted profiles.
The strong redshift and red wing enhancement were observed just after the moderate intensity with the blue wing enhancement.
This was the typical evolution of the Mg\emissiontype{II}~h~and~k profiles in the location from 171\farcs 0\ to 176\farcs 2\ in y$_{\rm slit}$.

As we have confirmed, the intensity, the redshift, and the line width in all the six lines had spatially and temporally good correlation with each other.
The moving flare kernel was observed simultaneously with the increase of \textit{Fermi} hard X-ray flux.
Therefore, the variation of three quantities would be involved in the variation of energy input by electron beams \citep{fang93}.
The maximum velocity of the redshift at $y_{\rm slit}=$172\farcs 4 was 62~km~s$^{-1}$ (Si\emissiontype{IV}), 79~km~s$^{-1}$ (C\emissiontype{II}), and 51~km~s$^{-1}$ (Mg\emissiontype{II}~h), 34~km~s$^{-1}$ (Ca\emissiontype{II}~K), 15~km~s$^{-1}$ (Ca\emissiontype{II}~8542~\AA), and 46~km~s$^{-1}$ (H$\alpha$) for each line.
These values were at the same order with the values in the previous reports \citep{ichimoto84,shoji95,liu15}.

Finally, we describe the evolution in the region of $176\farcs 2$\ to $178\farcs1$\ in $y_{\rm slit}$, which corresponds to the northern leading edge of the flare kernel in its decaying phase.
In contrast with the region that we reported in previous paragraphs, the blueshift in the Mg\emissiontype{II}~h line in this region was not followed by any redshift.
Instead, rather persistent blueshift was observed in the Mg\emissiontype{II}~h line (see figure~\ref{fig-tymap_iris}).
The blueshifted Mg\emissiontype{II}~h line profiles at $y_{\rm slit}>$176\farcs 2 had lower intensity than those at $y_{\rm slit}<$176\farcs 2.
In other lines, intensities in this region were small as well.
We discuss this long lasting blueshift in the subsection 5.6.

\begin{figure*}[htbp]
\begin{center}\FigureFile(160mm,1mm){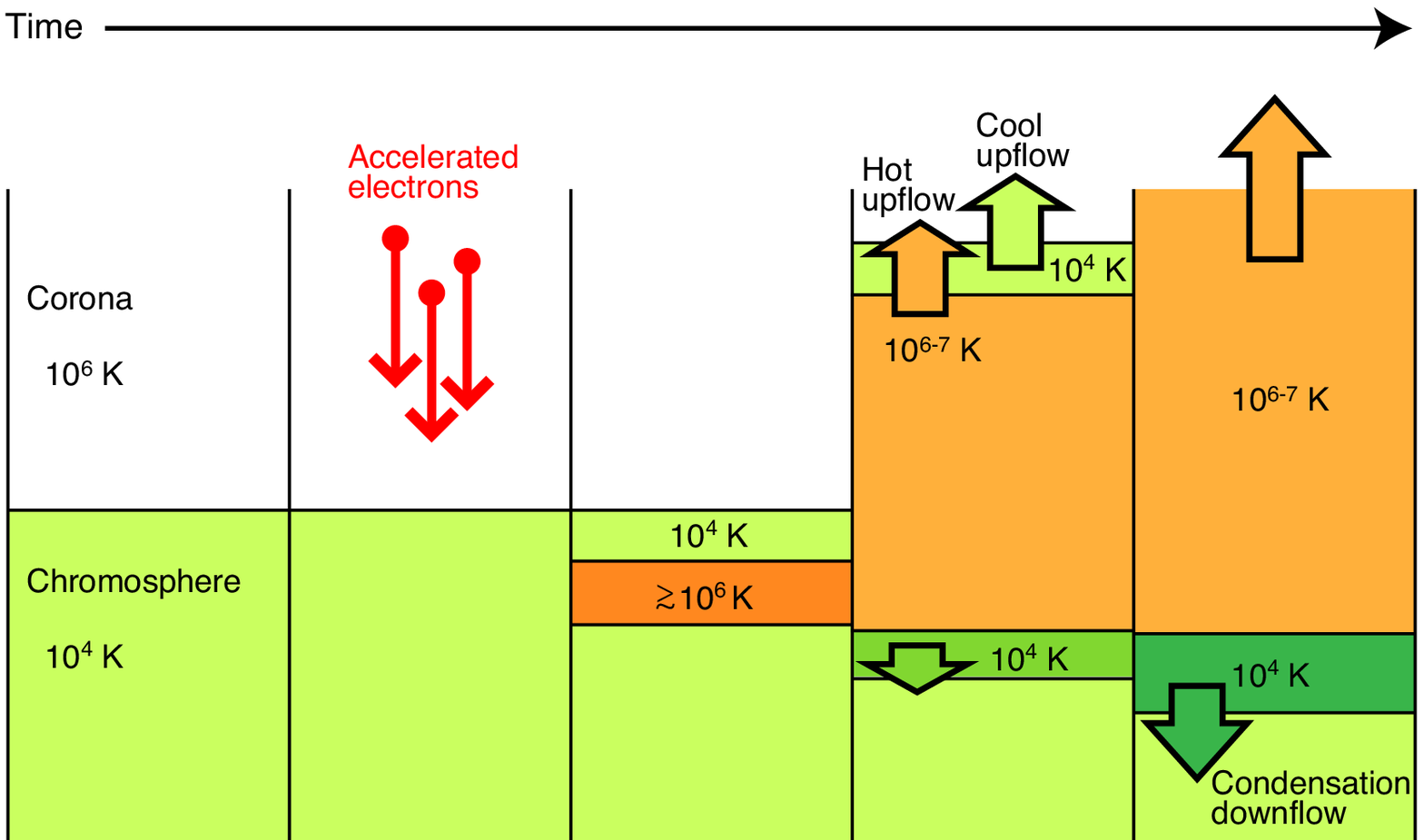}\end{center}
\caption{
A schematic cartoon of the one-dimensional cool upflow scenario that we discuss in this study.
The deep penetration of non-thermal electrons into the chromosphere causes an upflow of the chromospheric-temperature (cool) plasma lifted up by the expanding (hot) plasma, before the strong emission in the chromospheric lines from the downward moving condensation region.
(Color online)}
\label{fig-pon}
\end{figure*}

\section{Discussion}

\subsection{Observational results related to the blueshift in the Mg\emissiontype{II} lines}

We discovered the blue wing enhancement of the Mg\emissiontype{II}~h~and~k lines prior to the large intensity and redshift.
The blueshift in the wing amounts to 10.1~$\pm$ 2.6~km~s$^{-1}$ and it lasted for 9--48~s or for 28~s in average in the Mg\emissiontype{II}~h line.
It is also found that the blue-side peaks (h2v and k2v) were weaker than the red-side ones (h2r and k2r) in the Mg\emissiontype{II} h and k lines during this period. 
We discuss two scenarios for such blueshift: the cool downflow scenario in the sebsection~5.2 and the cool upflow scenario in the subsection 5.3, 5.4, and 5.5.
In addition, we observed the long-lasting blueshift that was not followed by any large redshift and intensity.
Note that we could not confirm any filament-like motion around the location where the blue wing enhancements were observed.

\subsection{Discussion of the blueshift: the cool downflow scenario}

First, let us think about a possibility that a downward motion of chromospheric plasma generates the blue asymmetry.
\citet{heinzel94} proposed a model in which the upper chromospheric layer moving downward absorbs the radiation of the red-side peak of emission in the H$\alpha$ line formed in deeper layer thus lowering its intensity in comparison to the blue one.
\citet{kuridze15} calculated the chromospheric line profiles and showed that the blue asymmetry can occur in the H$\alpha$ line, which is owing to the shift of maximum opacity of the overlaying layer to longer wavelengths.
However, in such cases, the intensity of the red-side peak is lower than that of the blue one, while in our observation the red-side peak was higher than the blue-side one in the Mg\emissiontype{II}~h~and~k lines.
Hence, this scenario cannot explain the observed feature of the Mg\emissiontype{II}~h~and~k lines.

\subsection{Discussion of the blueshift: the cool upflow scenario}

On the other hand, \citet{canfield90} discussed that an upward motion of chromospheric plasma can cause the blue asymmetry when high-energy electrons rush into the deep chromosphere in the early phase of the flare.
Although in their study the blueshift in the H$\alpha$ line occurred when the intensity level maximized, in our observation the blueshift in the Mg\emissiontype{II}~h line lasted for 9--48~s at each pixel before the largest intensity and redshift.
We present a scenario for our observation in figure~\ref{fig-pon}, which is a cartoon of temporal variation of plasma dynamics at a flare footpoint;
high-energy electrons penetrate into the deeper layer, and the plasma is heated to orders of magnitude of 10$^{6-7}$~K.
Then, the chromospheric cool plasma above the expanded plasma is lifted up, which can be observed as blueshift in chromospheric lines until the radiation from the condensation region become too strong.
This scenario is partially implied in one-dimensional hydrodynamic simulations of flaring loops by thick-target heating \citep{nagai84, allred05, kennedy15, rubio15b}.
In \citet{allred05}, the cool (${\rm Log}\ T [{\rm K}] =$ 4.5--4.7) and dense plasma is moving upward above the hot (${\rm Log}\ T [{\rm K}] =$ 6.0--6.5) plasma during the explosive phase of both moderate and strong heating cases, while the radiation from the chromospheric condensation region seems to be already strong in this phase.
In \citet{kennedy15}, the non-thermal electron beam heating causes a region of high-density cool (${\rm Log}\ T [{\rm K}] =$ 4.0--4.5) material propagating upward ($\sim$ 100~km~s$^{-1}$) through the corona during the explosive phase. 
Then the temperature of this material rapidly increased because the energy input by non-thermal electrons and thermal conduction was too large to be radiated away.
We suppose that in specific condition, flare heating can balance with radiative cooling in such cool dense plasma and the material can survive for longer time.
The blue wing enhancement in the Mg\emissiontype{II}~h line observed in our study may be attributed to such cool upflow above expanding hot plasma that is attributed to high energy electrons.
Actually, the blueshift in the Mg\emissiontype{II}~h line occurred in the second burst with the increase of the time derivative of the \textit{GOES} SXR flux and \textit{Fermi} HXR flux (figure~\ref{fig-lc_multi}), which is a proxy of the energy input by high energy electron beams.

\begin{figure*}[htbp]
\begin{center}\FigureFile(160mm,180mm){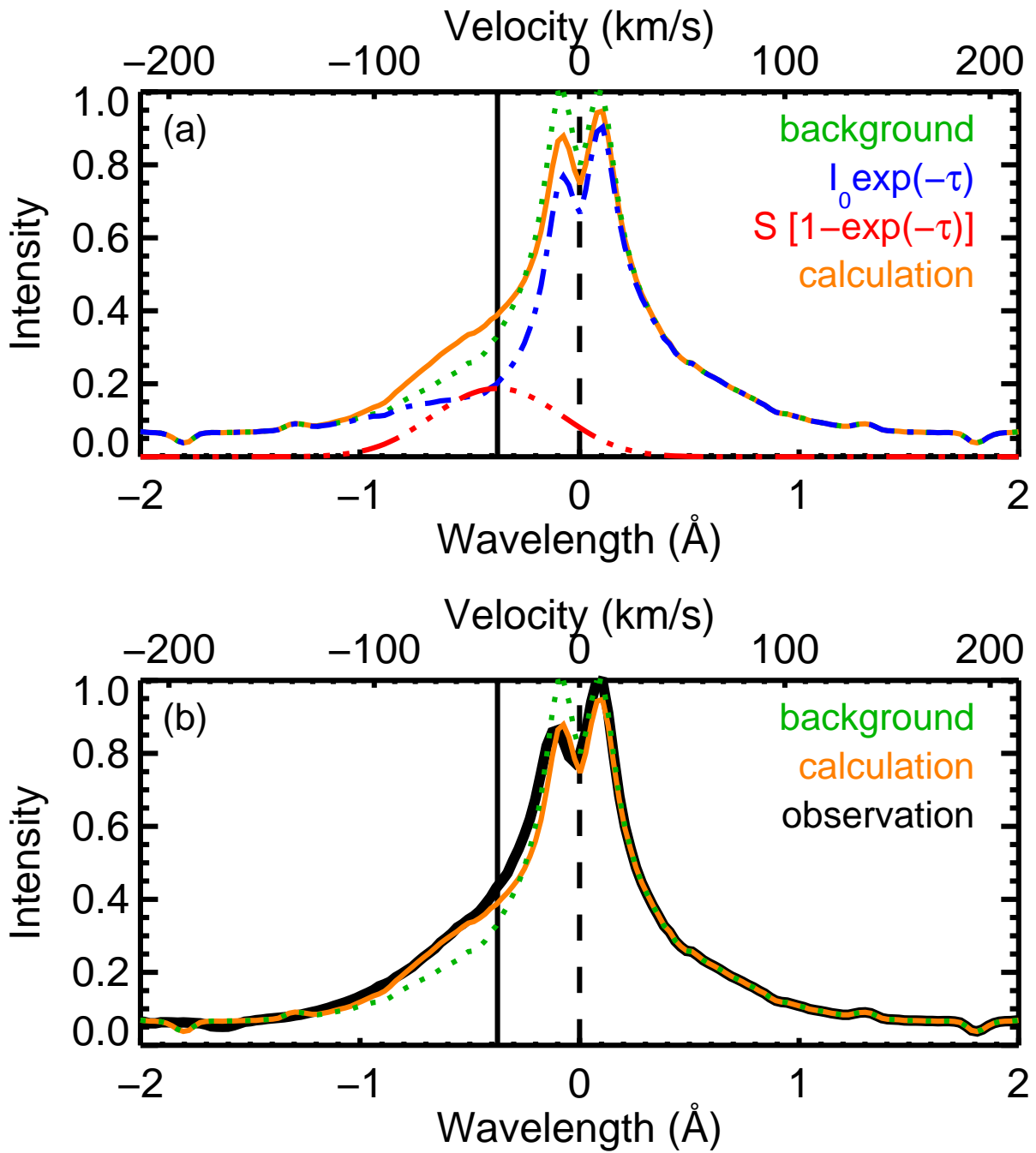}\end{center}
\caption{
(a) The cloud model calculation of the Mg\emissiontype{II}~h line profile with the values of free parameters: $V_{LOS}=-40$~km~s$^{-1}$, $V_{turb}=40$~km~s$^{-1}$, $\tau_0=0.5$, and $\alpha=0.6$.
The synthesized line profile well reproduces the averaged profile from the observation.
The calculated line profile $I(\Delta\lambda)$ is shown in the orange solid line, the background intensity $I_0(\Delta\lambda)$ in the green dotted line, the two terms in the radiative transfer equation $I_0{\rm exp}(-\tau(\Delta\lambda))$ and $S\left[1-{\rm exp}(-\tau(\Delta\lambda))\right]$ in the blue-colored dash-dotted line and red-colored dashed-three-dotted line, respectively.
Intensity is normalized by the background peak intensity $I_0(\Delta\lambda=0)$.
(b) The comparison of the calculated (orange thinner solid line, which is the same as shown in the panel (a)) and observed (black thicker solid line) profile of the Mg\emissiontype{II}~h line.
The background intensity $I_0(\Delta\lambda)$ is also drawn in the green dotted line.
The vertical dashed line shows the position of the rest wavelength corresponding to $\Delta\lambda=0$.
The solid vertical line indicates the wavelength position of the emission from the moving cloud.
Its line-of-sight velocity is $V_{LOS}=40$~km~s$^{-1}$.
Note that the bisector method generally leads to smaller shifts compared to cloud model fitting (The typical blueshift corresponded the velocity of 10.1~$\pm$~2.6~km~s$^{-1}$ by the bisector method in the observation).
(Color online)}
\label{fig-cloud}
\end{figure*}

\subsection{A cloud modeling based on the cool upflow scenario}

To discuss our result more quantitatively in terms of the cool upflow scenario, we calculated the Mg\emissiontype{II}~h line profile using a cloud model \citep{beckers64,tziotziou07}.
In this model, the equation of radiative transfer is expressed as
\begin{equation}
	I(\Delta\lambda) = I_0(\Delta\lambda)~e^{-\tau(\Delta\lambda)}+S\left[1-e^{-\tau(\Delta\lambda)}\right],
\label{eq1}
\end{equation}
where
\begin{equation}
	\tau(\Delta\lambda) = \tau_0~{\rm exp}\left[-\left(\frac{\Delta\lambda-\Delta\lambda_{LOS}}{\Delta\lambda_D}\right)^2\right],
\label{eq2}
\end{equation}
and
\begin{equation}
	\Delta\lambda_D = \frac{\lambda_0}{c}~\sqrt{\frac{2k_BT}{m_{Mg}}+V_{turb}^2}.
\label{eq3}
\end{equation}
$\Delta\lambda=\lambda-\lambda_0$ is the difference between the wavelength $\lambda$ and the rest wavelength of the Mg\emissiontype{II}~h line $\lambda_0$, 
$I(\Delta\lambda)$ is the flare intensity, 
$I_0(\Delta\lambda)$ is the background intensity, 
$\tau(\Delta\lambda)$ and $S$ are the optical thickness and averaged source function of the cloud, respectively, 
$\tau_0$ is the optical thickness of the cloud at the line center, 
$\Delta\lambda_{LOS}\equiv\lambda_0V_{LOS}/c$ is the shift of wavelength corresponding to the line-of-sight velocity of the cloud $V_{LOS}$ ($c$ is the light speed), 
$\Delta \lambda_D$ is the Doppler width, $k_B$ is the Boltzmann constant, 
$T$ is the temperature, 
$m_{Mg}$ is the atomic mass of Magnesium, and $V_{turb}$ is the turbulent velocity in the cloud.
In the calculation, we assume that the temperature of the cloud is $T$=10$^4$ K, and for the background intensity $I_0(\Delta\lambda)$ we use a symmetric line profile made from the red-side half of the averaged profile shown in the thick line in figure~\ref{fig-peak}(b).
In addition, $\alpha$ is defined as the ratio of the source function to the background intensity at the line center ($\alpha\equiv S/I_0(\Delta\lambda=0)$).
Figure~\ref{fig-cloud} shows a result of cloud model fitting where the values of free parameters are $V_{LOS}=-40$~km~s$^{-1}$, $V_{turb}=40$~km~s$^{-1}$, $\tau_0=0.5$, and $\alpha=0.6$.
The result produces a line profile similar to the averaged profile from the observation.
We stress that the observed peak difference (the blue-side peak is smaller than the red-side peak) is attributed to the attenuation by the cloud $I_0(\Delta\lambda)~{\rm exp}\left[-\tau(\Delta\lambda)\right]$ and the blue wing enhancement is due to the brightening from the cloud itself $S\left\{1-{\rm exp}\left[-\tau(\Delta\lambda)\right]\right\}$.
In this point, the scenario of upward moving cool plasma is consistent with our observation.
In addition, the line-of-sight velocity of the cloud $V_{LOS}$ in the cloud modeling is larger (40~km~s$^{-1}$) than the
the velocity by the bisector method in the observation (10.1~$\pm$~2.6~km~s$^{-1}$).
Note that the bisector method generally leads to smaller shifts compared to cloud model fitting.

It is true that we did not perform the full non-LTE modeling of Mg\emissiontype{II} line profiles, which can be done in a next paper.
Therefore, we do not exclude other possibilities.
However, the cloud model itself represents the formal solution of the non-LTE transfer equation assuming the simple model with constant quantities within the moving layer.
In comparison with the observed profiles, the cloud model provides a reasonable estimate of the flow velocity and the line width, along with other two quantities that are the line source function and the line center optical thickness.
While the emission from the moving cloud produces the blue wing enhancement but the absorption by the cloud causes the lowering of the blue-side peak, the detailed non-LTE modeling will be needed to compute other quantities.
This will then require the model of temperature and density structure.
Such models have been recently produced using the hydrodynamical code (Nakamura et al. in preparation) and preliminary non-LTE modeling based on selected time snapshots indeed leads to qualitatively the same Mg\emissiontype{II} profiles as observed (P. Heinzel, private communication).

\subsection{Consistency of the cool upflow scenario with observational results in various lines}

Next, why was the blueshift not observed in other lines, when it appeared in the Mg\emissiontype{II}~h line?
Concerning the DST lines, one possibility is that the region with the blue wing enhancement (spatial scale $\sim$1\arcsec) was not resolved by the ground-based DST observation because of the seeing effect ($\sim$2\arcsec).
The other possibility is due to the opacity differences; if we assume the VAL-C atmosphere (averaged quiet Sun model: \cite{val3}) as an initial model and pay attention to the 10,700 K plasma, then the opacity of the Mg\emissiontype{II}~h line is estimated to be about 14 times larger than that of the H$\alpha$ line.
In addition, the opacity of the Mg\emissiontype{II}~h line is known to be larger than that of the Ca\emissiontype{II}~K line and the Ca\emissiontype{II}~8542~\AA\ line \citep{leenaarts13}.
Therefore, it is possible that cool upflow is seen in the Mg\emissiontype{II}~h line but not in the Ca\emissiontype{II}~K, Ca\emissiontype{II}~8542~\AA, and H$\alpha$ lines.
On the other hand, the opacity of the C\emissiontype{II} 1335~\AA\ line is similar to that of the Mg\emissiontype{II}~h line \citep{rathore15}.
Indeed, the blue-side peak was typically smaller than the red-side peak in the C\emissiontype{II} 1335~\AA\ line but no blue wing enhancement was observed in our study.
This can be interpreted in this way: the difference of the peak intensities can be understood by the same model as that of the Mg\emissiontype{II}~h line.
However, the source function of the upward moving cloud might be too small to cause blue wing enhancement in the C\emissiontype{II} 1335~\AA.
Finally, the Si\emissiontype{IV}~1403~\AA\ line forms at transition region temperatures around 10$^{5}$~K and thus is optically thin at 10$^{4}$~K.
The upward moving cool cloud might be transparent in the Si\emissiontype{IV}~1403~\AA\ line.

\subsection{Interpretation of the long-lasting blueshift}

While the bluesift was followed by the redshift from 170\farcs 0 to 176\farcs 2 in $y_{\rm slit}$ in the Mg\emissiontype{II}~h line, the long-lasting blueshift without clear redshift was observed from 176\farcs 2 to 178\farcs 1 in $y_{\rm slit}$. This cannot be explained by the cool upflow scenario discussed above, where cool plasma is lifted up by the expanding hot plasma.
The blueshifted Mg\emissiontype{II}~h line profiles at $y_{\rm slit}>$176\farcs 2 had lower intensity than those at $y_{\rm slit}<$176\farcs 2.
That means the energy flux at $y_{\rm slit}>$176\farcs 2 was so small and heating was weak, which resulted in smaller kinetic energy of downward flow.
Thus, the downward velocity and downward compression would have been weak, and bright condensation downflow would not be observed.
On the other hand, it is reasonable that cool upflow is associated with such small energy flux.
There are possible interpretations for both the cases with the small energy input in the deeper and the middle or upper chromsphere.
The small energy input in the deeper chromosphere could lead to the local heating, causing gas pressure increase to generate a shock wave, which drives the cool upflow when it crosses the transition layer \citep{osterbrock61, suematsu82}.
In this case, the transition region moves upward, which results in cool upflow without any expanding hot plasma.
If the small energy input is given in the middle or upper chromosphere to increase the gas pressure to a few times of the initial value, then the gas pressure gradient force accelerates the cool plasma upward \citep{shibata82}.
Also in this case, the cool upflow is not associated with any expanding hot plasma.

\subsection{Summary and future work}

We performed coordinated observations of NOAA AR 12205, which produced the C-class flare on 2014 November 11, with the \textit{IRIS} and the DST at Hida Observatory.
Using spectral data in the Si\emissiontype{IV}~1403~\AA, C\emissiontype{II}~1335~\AA, and Mg\emissiontype{II}~h~and~k lines from \textit{IRIS} and the Ca\emissiontype{II}~K, Ca\emissiontype{II}~8542~\AA, and H$\alpha$ lines from DST, we investigated the temporal and spatial evolution around the flare kernel during the flare.
The flare kernel apparently moved along the \textit{IRIS} slit.
In the Mg\emissiontype{II}~h line, the leading edge of the flare kernel showed intensity enhancement in the blue wing, and small intensity of the blue-side peak (h2v) compared to the red-side one (h2r).
Then, the drastic change of the intensity in the red wing occurred.
The blueshift lasted for 9--48~s with a typical speed of 10.1~$\pm$~2.6~km~s$^{-1}$ and it was followed by strong redshift with a speed of up to 51~km~s$^{-1}$ detected in the Mg\emissiontype{II}~h line.
The strong redshift was a common property for all six spectral lines used in this study but the blueshift prior to it was found only in the Mg\emissiontype{II} lines.
A cloud modeling of the Mg\emissiontype{II}~h line suggests that the blue wing enhancement with such peak difference can be caused by a chromospheric-temperature (cool) upflow.
We discuss a scenario in which an upflow of cool plasma is lifted up by expanding (hot) plasma owing to the deep penetration of non-thermal electrons into the chromosphere.
In addition, the long-lasting blueshift was also observed in the Mg\emissiontype{II}~h line, which was not followed by any large redshift and intensity enhancement.
This occurred at the northern leading edge of the flare kernel in its decaying phase.
Such long-lasting blueshift can be explained by cool upflow caused by so small energy flux into the lower atmosphere that expanding hot upflow is not produced.

The chromospheric lines used in this study are formed under the non-LTE condition.
The observed line profile is influenced by the three-dimensional radiative field depending on the temperature and density structure.
In the case of no velocity field, the intensities in the line core and wings reflect the information about the higher and lower levels of the atmosphere, respectively, since the absorption coefficient has its maximum in the line core and is smaller in the wings.
In addition, there are large velocity gradients during solar flares.
Therefore, it is not so simple to diagnose the velocity field at a certain height from spectral line profiles.
Modeling of the flaring atmosphere and synthesizing chromospheric line profiles has been conducted so far (e.g., \cite{nagai84,allred05,rubio15a},\ \yearcite{rubio15b},\ \yearcite{rubio16},\ \cite{rubio17}).
Our result about the blueshift in the Mg\emissiontype{II} lines will restrict the further modeling.
There would be both cases of flaring regions with and without a signature of blueshift in the chromospheric lines.
Even at a single location, the chromospheric response to flares differs for different observing lines.
The initial condition and energy input may affect the evolution of cool upflow significantly 
and the spectra would also change sensitively by the surrounding environment.
More studies are needed in the future to clarify what causes these differences and to understand the heating mechanism of the flaring chromosphere from the spectroscopic observations.

\begin{ack}
A.T. would like to thank Dr. Wei Liu and Dr. Fatima Rubio da Costa for helpful discussion.
\textit{IRIS} is a NASA small explorer mission developed and operated by LMSAL
with mission operations executed at NASA Ames Research center and major contributions
to downlink communications funded by ESA and the Norwegian Space Centre.
\textit{SDO} is part of NASA's Living With a Star Program.
This work was partly carried out on the Solar Data Analysis System operated
by the Astronomy Data Center in cooperation with the Hinode Science Center of NAOJ.
This work was supported by JSPS KAKENHI Grant Numbers, JP17J07733 (PI: A.T.), JP16K17663 (PI: T.J.O.), JP17K14314 (PI: T.K.), JP15K17772 (PI: A.A.), JP15H05814 (PI: K.I), JP16H03955 (PI: K.S.), and JP25220703 (PI: S. Tsuneta).
PH acknowledges support from the Czech Funding Agency through the grant No. 16-18495S and from the Kyoto University during his stay in Japan.

\end{ack}


\end{document}